\documentclass[aps,prl,twocolumn,showpacs,superscriptaddress]{revtex4-1}
\usepackage[latin1]{inputenc}
\usepackage{epstopdf}
\usepackage{graphicx}
\usepackage{indentfirst}
\usepackage{amsmath, amsthm, amssymb} 
\usepackage{bm}
\usepackage{epsfig}
\input epsf
\usepackage{float}
\usepackage[colorlinks=true, a4paper=true, pdfstartview=FitV, linkcolor=blue, citecolor=blue, urlcolor=blue]{hyperref}
\usepackage{epsfig}
\usepackage{braket}

\begin{document}

\title{Bump-on-tail instability of twisted excitations in rotating cold atomic clouds} 

\author{J. D. Rodrigues}
\affiliation{Instituto de Plasmas e Fus\~{a}o Nuclear, Instituto Superior T\'{e}cnico, Universidade de Lisboa, 1049-001 Lisboa, Portugal}

\author{H. Ter\c cas}
\affiliation{Institute f\"ur Theoretische Physik, Universit\"at Innsbruck, 6020 Innsbruck, Austria}
\affiliation{Institute for Quantum Optics and Quantum Information, Austrian Academy of Sciences, 6020 Innsbruck, Austria}
\affiliation{Physics of Information Group, Instituto de Telecomunica\c{c}\~oes, Lisbon, Portugal}

\author{J. T. Mendon\c ca}
\affiliation{Instituto de Plasmas e Fus\~{a}o Nuclear, Instituto Superior T\'{e}cnico, Universidade de Lisboa, 1049-001 Lisboa, Portugal}


\begin{abstract}
\begin{footnotesize}

We develop a kinetic theory for twisted density waves (phonons), carrying a finite amount of orbital angular momentum, in large magneto optical traps, where the collective processes due to the exchange of scattered photons are considered. Explicit expressions for the dispersion relation and for the kinetic (Landau) damping are derived and contributions from the orbital angular momentum are discussed. We show that for rotating clouds, exhibiting ring-shaped structures, phonons carrying orbital angular momentum can cross the instability threshold and grow out of noise, while the usual plane wave solutions are kinetically damped.

\end{footnotesize}
\end{abstract}

\maketitle

\par

{\it Introduction}. It is an established fact that electromagnetic radiation can carry a finite amount of angular momentum, due to the resulting contributions from the spin and orbital angular momentum degrees of freedom. Although the spin is always an intrinsic variable, orbital angular momentum (OAM) may be either intrinsic or extrinsic \cite{intro1}. Spin to orbital angular momentum conversion may occur in an homogeneous and isotropic medium \cite{intro2}. More recently, attention has been given to the exchange of orbital angular momentum between EM radiation and different systems, such as electrostatic waves in plasmas \cite{tito_paraxial}, nonlinear media \cite{intro3}, acoustic vortices in optical fibers \cite{intro5} or even sub-meter sized metallic objects \cite{intro6}. Nevertheless, OAM is not a feature exhibited only by light. In fact, vortex phonons have also been proposed to manipulate particles and small objects \cite{intro7, intro8}. In magnetic crystals, for example, spin-orbit coupling can promote the excitation of phonons carrying a finite amount of orbital angular momentum \cite{intro9}. 

\par
The manipulation of OAM became attractive in cold atomic systems as well since the observation of optical pumping of OAM in Cs atoms \cite{tabosa} and the vortex nucleation in Bose-Einstein condensates \cite{andersen}, by means of Laguerre-Gauss (LG) beams. Cold atoms confined in large magneto-optical traps are particularly interesting, as the atomic dynamics is significantly altered due to the appearance of a collective interaction mediated by the multiple scattering of light \cite{sesko_2, dalibard}. In this regime, the atoms experience a Coulomb-like long-range interaction \cite{Q}, allowing for an analogy with a one component trapped plasma \cite{livro}.\par  
In this paper, we theoretically describe a combined effect of OAM and the multiple-scattering of light: the onset of a dynamical instability in rotating clouds of cold atoms triggered by the OAM of the hybrid phonons - the elementary excitations in large magneto-optical traps (MOT). Hybrid phonons, with properties similar to both plasma and acoustic waves \cite{tito_2008}, have been at the origin of a large set of exotic phenomena such as phonon-lasing \cite{phonon_laser}, classical rotons \cite{rotons}, photon bubbles \cite{bubbles} or the dynamical Casimir effect \cite{Casimir}. 
\begin{figure}[t!]
\includegraphics[scale=1]{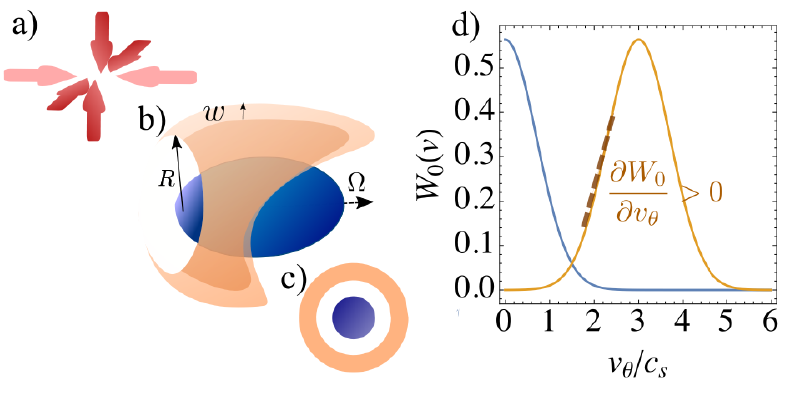}
\caption{(color online) Illustration of the experimental setup. (a) The transverse beams (darker red), which are slightly misaligned, are of much higher intensity than the beams along the longitudinal direction (lighter red). (b) Perspective and (c) top views of the density profile for a rotating trap. The central cloud (blue) is surrounded by atoms located at the shell of mean radius $R$ and width $w$ rotate at the angular velocity $\Omega$, as a consequence of the misalignment of the trapping laser along the transverse direction. (d) Poloidal velocity distribution function. The speed distribution of the atoms of the rotating shell contains a positive derivative (bump-in-tail), which will originate the instability.}
\label{fig_scheme}
\end{figure}
In our setup, and contrary to what happens with light - where the phase singularities are imposed by phase modulation - the collective excitations (phonons) in cold atoms acquire OAM due to the rotation of the cloud, as a consequence of the total angular momentum conservation. \par The experimental setup we have in mind consists of a large laser-cooled $^{87}$Rb gas, with a typical total number of atoms $N\sim 10^{9}-10^{11}$.  Via a slight misalignment of the trapping laser beams along, as described in the experiments by Sesko et al. \cite{sesko_2}, we produce a Saturn-like structure (see Fig. \ref{fig_scheme}). The external clump of atoms then rotates at a constant angular velocity $\Omega$ of the order the thermal velocity of the effective plasma frequency $\omega_p$ \cite{tito_2008}. The modification we require in respect to usual experimental configurations is that the laser beam intensity along the longitudinal direction should be much smaller than the transverse ones, such that cloud is cigar shaped. As a result, the equilibrium of cooling and trapping produce the configuration of a cigar-shaped Saturn-like structure as illustrated in Fig. \ref{fig_scheme}. The rotation of the exterior shell transfer angular momentum to the excitations (phonon) on top of the equilibrium configuration. We therefore show that such ``twisted" phonons lead to a kinetic instability, a type-$I_o$ oscillatory instability \cite{cross}, in a rotating cloud of cold atomic trap operating in the multiple-scattering regime. This instability is formally very close to the {\it bump-on-tail instability} observed in fluids and plasmas \cite{bittencourt, chen}, as the rotation induces an offset in the poloidal velocity distribution, allowing for the instability to develop. The important difference with respect to other system stems in the fact that only twisted phonons can trigger the instability in our case, while zero OAM excitations are usually enough for the bump-on-tail sort of kinetic instabilities.


\par
{\it Kinetic description}. The complete kinetics of cold atoms in magneto-optical traps is described by a Vlasov-Poisson equation of the form 
\begin{equation}
\label{eq:vlasov}
\left( \frac{\partial}{\partial t} + \bm{v} \cdot \bm{\nabla} + \frac{\bold{F}}{m} \cdot \bm{\nabla}_{v} \right) W \left( \bold{r}, \bold{v}, t \right) = \mathcal{I}[W(\mathbf{r},\mathbf{v},t)] 
\end{equation}
where $W \left( \bold{r}, \bold{v}, t \right)$ is the distribution function normalized to the total number of particles $N = \int d \bold{r} \int d \bold{v} W \left( \bold{r}, \bold{v}, t \right)$. The total force $\mathbf{F}=\mathbf{F}_{\rm trap}+\mathbf{F}_{c}$ accounts for the cooling and trapping terms, $\mathbf{F}_{\rm trap}\simeq -\kappa \mathbf{r}-\alpha \mathbf{v}$, with $\kappa$ and $\alpha$ being the spring constant and the cooling rate, respectively, and the collective force $\mathbf{F}_c=-\bm \nabla V_c$ \cite{Q, tito_2008, livro, hugo_2013} with
\begin{equation}
\nabla^2 \bold{V}_c \left( \bold{r} , t \right) = - Q \int d \bold{v} W \left( \bold{r}, \bold{v}, t \right)
\label{eq:poisson}
\end{equation}
where $Q=\sigma_L \left(\sigma_R - \sigma_L \right) I/c$ represents an effective charge of the atoms, with $\sigma_R$ and $\sigma_L$ the scattering and absorption cross sections \cite{dalibard, sesko_1, sesko_2, pruvost}, respectively, and $I$ the light intensity. The term $\mathcal{I[W]}$ in Eq. (\ref{eq:vlasov}) represents the collision integral and describes the source of hot reservoir atoms and the collisions between them \cite{romain}. As we are interested in small deviations from equilibrium only (thus ruling out far-from-equilibrium effects) which occur at much slower time-scales that the dynamics of hot atoms, we may safely neglect the contribution of the collision integral. Moreover, for large traps (typically with $N \sim 10^9 - 10^{10}$), and appropriate laser detuning, the density profiles are approximately constant \cite{scaling_1, scaling_2}, or at least slowly varying at the scale of the fluctuations we are about to describe. As such, we should invoke the local-density approximation (LDA) which amounts to neglect the trapping inhomogeneities. \par
In order to account for small amplitude fluctuations around the LDA equilibrium $W_0$, we linearize Eq. (\ref{eq:vlasov}) and (\ref{eq:poisson}), with $W = W_0 + \delta W$ and $V_c = V_{c,0} + \delta V_c$, yielding
\begin{equation}\label{eq:vlasov_lin}
\left( \frac{\partial}{\partial t} + \bm{v} \cdot \bm{\nabla} \right) \delta W =  \frac{\bold{\delta F}_c}{m} \cdot \bm{\nabla}_{v} W_0,\quad {\rm and}
\end{equation}
\begin{equation}\label{eq:poisson_lin}
\quad  \nabla^2 \delta V_c = - Q \int d \bold{v} ~\delta W \left( \bold{v} \right).
\end{equation}
In what follows, we consider a cigar-shaped trap elongated along the $z-$direction, in the absence of rotation (see Fig. \ref{fig_scheme}). Experimentally, this aspect ratio is obtained by decreasing the intensity of the trapping lasers along the $z$-direction. In this case, a slowly-varying amplitude density fluctuation propagating along the $z-$axis as $e^{ikz}$ must be solution of the following equation
\begin{equation}\label{eq:envelope}
\nabla^2 \delta V_c \simeq \left( \nabla_\perp^2 - k^2 +2ik \frac{\partial}{\partial z} \right) \delta V_c.
\end{equation}
Under this approximation, the potential $\delta V_c$ also satisfies the paraxial equation $\left( \nabla_\perp^2 +2ik \frac{\partial}{\partial z} \right) \delta V = 0$ \cite{tito_paraxial}. Such a condition reduces Poisson's Eq. (\ref{eq:poisson_lin}) to $k^2 \delta V_c= Q \int d \bold{v} \delta W\left( \bold{v} \right)$. The solution to the paraxial equation can be given as a linear combination of Laguerre-Gauss (LG) modes as
\begin{equation}
\label{eq:LG1}
\delta V_c \left( \bold{r}, t \right) = \sum_{pl} \delta V_{pl} F_{pl} \left( r, z \right) e^{i l \theta} e^{ikz-i \omega t} 
\end{equation}
with $\delta V_{pl}$ being the mode amplitude (assumed to be small) and the integers $p$ and $l$ the radial and azimuthal mode numbers, respectively. The orthogonality of the LG modes reads $\Braket{F_{pl} \mid F_{p'l'}} = \delta_{pp'} \delta_{ll'}$, with the mode functions $F_{pl} \left(r,z \right)$ given by
\begin{equation}\label{eq:LG2}
F_{pl} \left( r,z \right) = C_{pl} \zeta^{\mid l \mid} L_p^{\mid l \mid} e^{-\zeta/2}
\end{equation} 
with $\zeta \equiv r^2/w^2(z)$ and $w(z)$ the wave beam waist. The normalization constants $C_{pl}$ are given by $C_{pl} = \frac{1}{2\sqrt{\pi}} \left[ \frac{\left(l+p \right)!}{p!} \right]^{1/2}$, with the usual Laguerre polynomials, $L_p^l \left( \zeta \right) = \frac{e^{\zeta}}{p! \zeta^{l}} \frac{d^p}{d\zeta^p} \left[ \zeta^{l+p} e^{-\zeta} \right]$.
\par

The fluctuations of the collective force are determined by $\bold{\delta F}_c = -i \bold{k}_{\rm eff} \delta V_c$ where the effective (paraxial) wave vector $\bold{k}_{\rm eff}$ reads \cite{tito_twisted}
\begin{equation}
\label{eq:k_eff}
\bold{k}_{\rm eff} = - \frac{i}{F_{pl}} \frac{\partial F_{pl}}{\partial r} \bold{e_r} + \frac{l}{r} \bold{e_{\theta}} + \left( k - \frac{i}{F_{pl}} \frac{\partial F_{pl}}{\partial z} \right) \bold{e_z}.
\end{equation}
Similarly, the fluctuations $\delta W$ can also be decomposed into a superposition of LG modes as
$\delta W \left( \bold{v} \right) = \sum_{pl} \delta W_{pl} \left( \bold{v} \right) F_{pl} \left( r, z \right) e^{i l \theta} e^{ikz-i \omega t}.$ By plugging in Eq. (\ref{eq:vlasov_lin}), multiplying by $F_{pl}$ and making use of the orthogonality relations, we obtain 
\begin{equation}\label{eq:mid1}
\delta W_{pl} = \frac{\delta \bold{F}_{c,pl}}{m \left( a+ib \right)}\cdot \bm{\nabla_v} W_0, 
\end{equation}
with $\delta \bold{F}_{c,pl} = - \bold{q}_{\rm eff} \delta V_{pl}$ and $\bold{q}_{\rm eff} = -i q_r \bold{e_r} + lq_{\theta} \bold{e_{\theta}} + \left( k - iq_z \right) \bold{e_z}$. Here, we set 
\begin{equation}\label{eq:mid2}
q_j = \int_0^{\infty} F_{pl} \frac{\partial F_{pl}}{\partial j} rdr, \qquad q_{\theta} = \int_0^{\infty} F_{pl}^2 dr
\end{equation}
with $j = \left\{ r,z \right\}$. The real quantities $a$ and $b$ are defined as $a = \left( \omega - kv_z \right) - lq_{\theta} v_{\theta}$ and $b=\left( q_r v_r + q_z v_z \right)$. Notice that no mixture between the model $p$ and $l$ is expected at the linear level, which is the sufficient description to understand the nature of low lying excitations. From Eq. (\ref{eq:mid1}) and the decomposition of $\delta W$ in LG modes, and noting that $a+ib = \left( \omega - \bold{q}_{\rm eff} \cdot \bold{v} \right)$, we finally derive the kinetic dispersion relation
\begin{equation}\label{eq:dispersion_ini}
1 - \frac{\omega_p^2}{k^2} \int \frac{\bold{q}_{\rm eff} \cdot \bm{\nabla}_v \tilde W_0}{\left(\omega - \bold{q}_{\rm eff} \cdot \bold{v} \right)} d\bold{v} \equiv 1+\chi (\omega)= 0,
\end{equation}
where we have defined the quantity $\tilde W_0(\mathbf{v}) = W_0(\mathbf{v},\mathbf{r}) / n_0(\mathbf{r})$ as normalized to the LDA density $n_0(\mathbf{r})$, and the effective plasma frequency $\omega_p^2 = Q n_0 / m$ \citep{tito_2008, livro}. This kinetic dispersion relation encodes the dynamics of twisted ($l\neq 0$) excitations above the equilibrium. Depending on the details of the velocity distribution $\tilde W(\mathbf{v})$, the OAM momentum may change the stability conditions of the system, as shown below.
\par

{\it Twisted Landau damping.} Assuming that $\left| q_r \right|, \left| q_z \right| \ll \left| l q_{\theta} \right|$, and separating the susceptibility into its real and imaginary part $\chi(\omega) = \chi'(\omega) + i \chi''(\omega)$, we rewrite Eq. (\ref{eq:dispersion_ini}) in the more explicit way
\begin{equation}\label{eq:suscpt_real}
\chi' \left( \omega \right) =  \frac{\omega_p^2}{k^2} \left[ \mathcal{P} \int \frac{\partial W_0 / \partial v_z}{\left( v_z - u_z \right)} d \bold{v} + \mathcal{P} \int \frac{\partial W_0 / \partial v_{\theta}}{\left( v_{\theta} - u_{\theta} \right)} d \bold{v} \right],
\end{equation}
where $\mathcal{P}$ denotes the Cauchy principal part of the integral. The two poles in the integration occur respectively for $u_z = \left( \omega - l q_{\theta} v_{\theta} \right) / k$ and $u_{\theta} = \left( \omega - k v_z \right) / lq_{\theta}$. On the other hand, for the imaginary part of the susceptibility we can write
\begin{equation}\label{eq:suscpt_imaginary}
\chi'' \left( \omega \right) =  - \pi \frac{\omega_p^2}{k^2} \left[ \left(\frac{\partial W_0}{\partial v_z} \right)_{v_z=u_z} + \left(\frac{\partial W_0}{\partial v_{\theta}} \right)_{v_{\theta}=u_{\theta}}  \right].
\end{equation}
For low enough temperatures, it is plausible to assume excitations with a phase speed much larger than the width of the distribution, $u_z \gg v_z$ and $u_{\theta} \gg v_{\theta}$, which allows us to expand the denominator in Eq. (\ref{eq:suscpt_real}) as
\begin{equation}
\mathcal{P} \int \frac{\partial W_0 / \partial v_j}{\left( v_j - u_j \right)} d \bold{v} \simeq \frac{1}{u_j^2} \int W_0 \left( v_j \right) \left( 1+3 \frac{v_j^2}{u_j^2} \right) d \bold{v},
\end{equation}
with $j=\left\{z, \theta \right\}$. The real part of the mode frequencies can be readily obtained by setting $\chi'\left( \omega \right) = 0$, which read
\begin{equation}\label{eq:dispersion}
\omega^2 = \omega_p^2  \left(1+ \frac{1}{\eta^2} \right) + 3 c_s^2 k^2, 
\end{equation}
where we have defined the speed of sound as $c_s^2 = \Braket{v^2} = \Braket{v_z^2} + \Braket{v_{\theta}^2} \frac{1}{\eta^4}$ with $\Braket{v_j^2} = \int v_j^2 W_0 \left( \bold{v} \right) d \bold{v}$ and $j=\left\{z, \theta \right\}$. The paraxial parameter $\eta$, accounting for the OAM of the excitation, is given by $\eta = k / l q_{\theta}$. Assuming that axial propagation is dominant, $\eta > 1$ - where the paraxial approximation makes sense - we get a small correction to the hybrid phonon modes derived in \cite{tito_2008}.

\par
\begin{figure}
\centering
\includegraphics[width=1\linewidth]{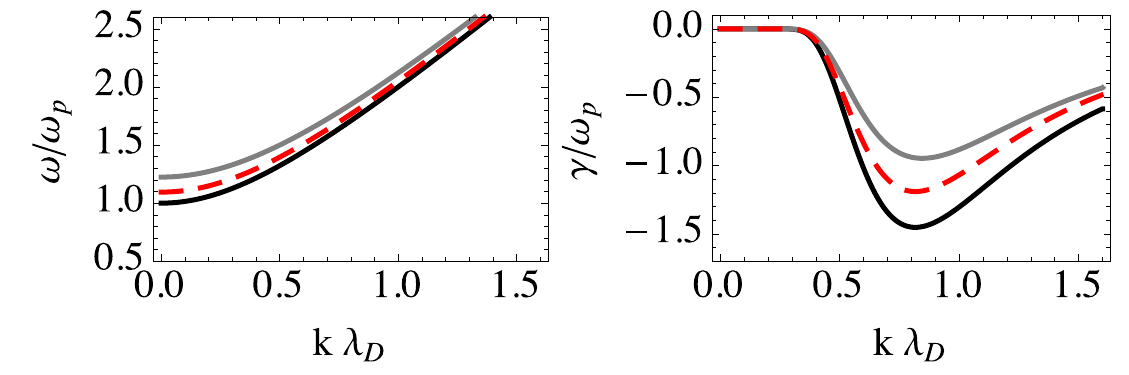}
\caption{(color online) Real (left panel) and imaginary (right panel) part of the dispersion relation for $\eta^2 = 2$ (gray line) and $\eta^2 = 5$ (red dashed line). The usual phonon dispersion (corresponding to the plane wave limit $\eta^2 \rightarrow \infty$) \cite{tito_2008, livro} (black solid line) is presented here for comparison.}
\label{fig:disp}
\end{figure}

Let us now examine the kinetic (Landau) damping of the twisted excitations. For that purpose we go back to Eq. (\ref{eq:suscpt_imaginary}) and note that the damping rate, $\gamma$, is given by $\gamma = \frac{\chi'' \left(\omega \right)}{\left( \partial \chi' / \partial \omega \right)} \simeq \omega \chi'' \left( \omega \right)$. The overall damping rate is the sum of two contributions, $\gamma = \gamma_z + \gamma_{\theta}$, from the longitudinal and perpendicular subspaces, respectively. We can now assume a Maxwellian distribution of the form $\tilde W_{0,j} \left( v_j \right) = 1 / c_s \sqrt{\pi} e^{-v_j^2 / c_s^2}$, with $j= \left\{ z, \theta \right\}$ and finally obtain
\begin{equation}\label{eq:damping}
\gamma = - \frac{2 \sqrt{\pi} \omega_p}{\left( k \lambda_D \right)^3} \left[ e^{-1/\left(k \lambda_D \right)^2} e^{-1/ \eta^2} + \eta e^{-\eta^2 / \left( k \lambda_D \right)^2} e^{-\eta^2}   \right],
\end{equation}
with $\lambda_D = c_s / \omega_p$ being an analogue of the Debye length, denoting the typical size of the hybrid phonons in the trap. In the limit of zero OAM, $\eta^2 \rightarrow \infty$, we recover the expected result $\gamma = - \frac{2 \sqrt{\pi} \omega_p}{\left( k \lambda_D \right)^3}  e^{-1/\left(k \lambda_D \right)^2}$. These results, as well as the dispersion relation given by Eq. (\ref{eq:dispersion}) are illustrated in Fig. (\ref{fig:disp}). 
As we can observe, $\gamma$ is negative for all modes, which corresponds to the Landau damping mechanism. However, excitations carrying OAM are shown to be less damped than the plane wave solutions. In what follows we shall demonstrate that, in the case of a rotating cloud of cold atoms, and for a particular range of achievable experimental conditions, the vortex phonons can cross the instability threshold and grow out of noise while the plane wave solutions remain kinetically damped.

 \par
\begin{figure}
\centering
\includegraphics[width=1\linewidth]{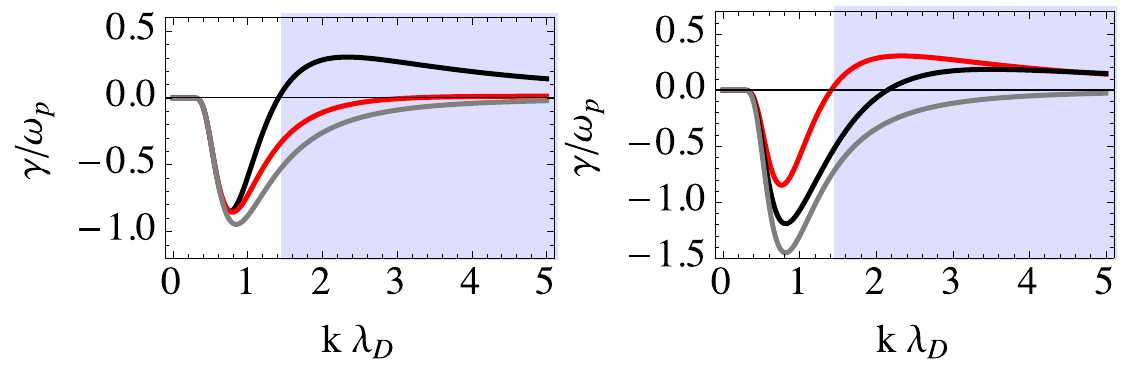}
\caption{(color online) Left panel: Kinetic damping rate for $\eta^2 = 2$ and different poloidal velocities, $\beta = 2$ (black line) and $\beta = 3$ (red line). The shaded region corresponds to an instability. The gray line corresponds to the non-rotating cloud, $\beta=0$. (right panel) Kinetic damping rate for constant poloidal velocity, $\beta=2$, and different paraxial parameters, $\eta^2 = 2$ (red line) and $\eta^2 = 5$ (black line). The plane wave solution, $\eta^2 \rightarrow \infty$ corresponds to the gray line.}
\label{fig:inst}
\end{figure}
\begin{figure}
\centering
\includegraphics[width=1\linewidth]{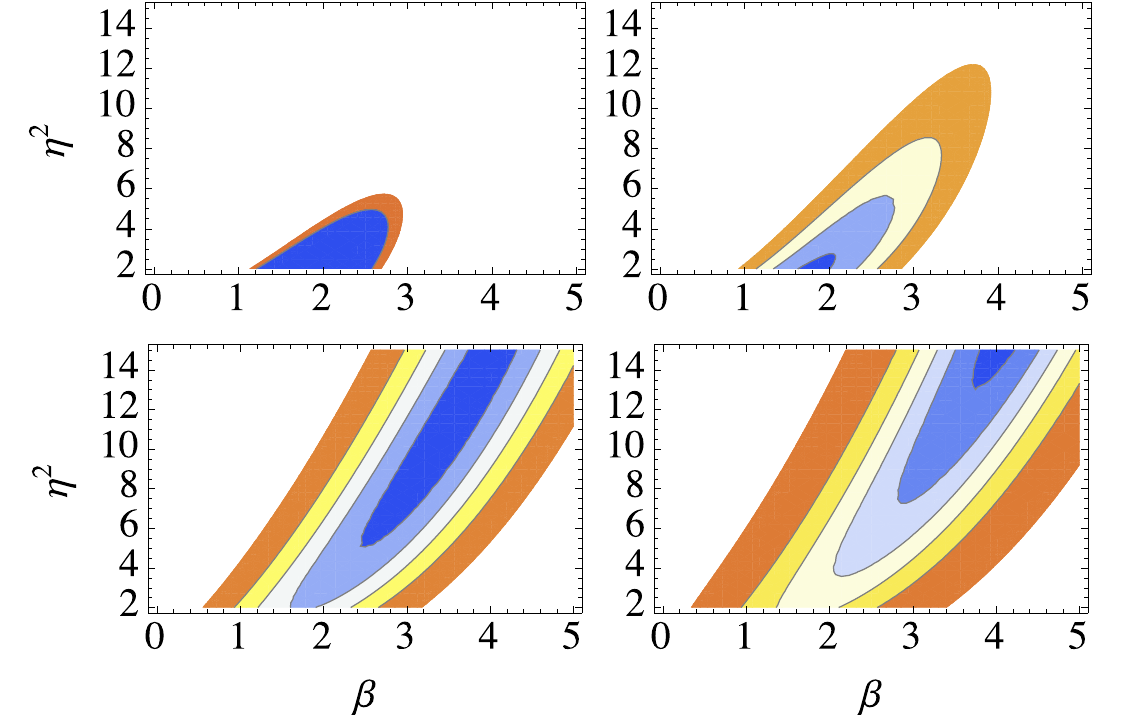}
\caption{(color online) Stability diagram for vortex phonons in the $\left( \beta, \eta^2 \right)$ plane for $k \lambda_D = 2$ (top left), $k \lambda_D = 2.5$ (top right), $k \lambda_D = 5$ (bottom left) and $k \lambda_D = 10$ (bottom right). The coloured region corresponds to the range of parameters for which a modulational instability occurs. Blue regions correspond to higher growth rates.}
\label{fig:diag1}
\end{figure}
\begin{figure}
\centering
\includegraphics[width=1\linewidth]{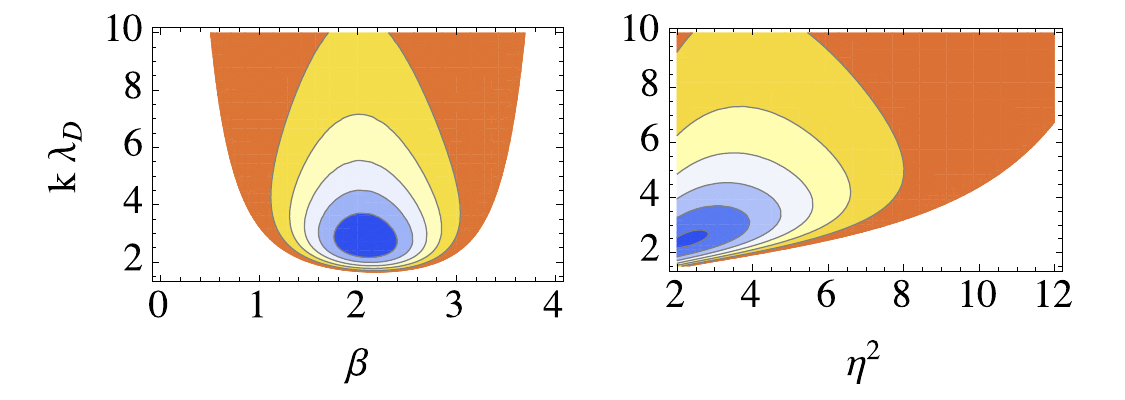}
\caption{(color online) Stability diagram for vortex phonons in the $\left( \beta, k \lambda_D \right)$ plane for $\eta^2 = 3$ (left panel) and in the $\left( \eta^2, k \lambda_D \right)$ plane for $\beta = 2$ (right panel).}
\label{fig:diag2}
\end{figure}

{\it Bump-on-tail instability.} Rotating clouds of cold atoms, resulting from small misalignments in the trapping beams, often present ring-shaped density profiles \cite{sesko_2, ring_1, ring_2, ring_3}, which are well described by a radial density function $n \left( r \right) = n_0 e^{-\left(r- R \right)^2 / w^2}$, where $R$ is the ring radius and $w$ its width (see Fig. \ref{fig_scheme}). In this case, the system can be approximately described by a rotating rigid body, $v_\theta \left( r \right) = \Omega r$ \cite{sesko_2}, where $\Omega$ is the angular frequency of rotation. By plugging this condition in the previous density function, we can derive a normalized poloidal velocity distribution function
\begin{equation}
 \tilde W_0 \left( v_\theta \right) = \frac{1}{\Delta c_\theta \sqrt{\pi}} e^{- \left(v_\theta - \Omega R \right)^2 / \Delta c_\theta^2},
 \end{equation}
with $ \Delta c_\theta = w \Omega$ denoting the width of the distribution. The interpretation of the later as a distribution function is justified by the typical scales of the experiments \cite{sesko_2}. For a rotating frequency of $\Omega \sim (2 \pi) 100$ Hz and a ring width $w \sim 0.5$ mm, we get $\Delta c_\theta = w \Omega \sim 0.3$ m/s. Such a velocity distribution width is of the same order of the thermal velocity of a $^{87}$Rb MOT at $T=200$ $\mu$K ($c_s \sim 0.25$ m/s). A typical ring radius of $R \sim 2$ mm corresponds to a poloidal velocity of $\Omega R \sim 1$ m/s. Plugging the new poloidal velocity distribution in Eq. (\ref{eq:suscpt_imaginary}), and noting that we can approximate $\lambda_D = c_s / \omega_p \simeq \Delta c_\theta / \omega_p$, which significantly simplifies the calculation, yields a new expression for the kinetic damping term
\begin{eqnarray}\nonumber
\gamma_\Omega= && - \frac{2 \sqrt{\pi} \omega_p}{\left( k \lambda_D \right)^3} \left[ e^{-1/\left(k \lambda_D \right)^2} e^{-1/ \eta^2} + \right.\\
&& \left. \eta \newline \left( 1 - \frac{\beta}{ \eta} \left( k \lambda_D \right) \right) e^{-\eta^2 / \left( k \lambda_D \right)^2} e^{-\eta^2 \left(1-\beta / \eta \right)^2} \right],
\end{eqnarray}
with $\beta = \Omega / c_s$. Note that, for $\beta = 0$, which corresponds to a regular Maxwell-Boltzmann distribution for non rotating clouds, we recover the previous result of Eq. (\ref{eq:damping}), as expected, where small oscillations in the homogeneous solution are always damped. These results are illustrated in Fig. \ref{fig:inst} and stability diagrams in terms of the appropriate parameters, $\eta^2$ and $\beta$, are sketched in Fig. \ref{fig:diag1} and Fig. \ref{fig:diag2}. A type $I_o$ $\left( k \neq 0, \omega \neq 0 \right)$ \cite{cross} oscillatory modulational instability can then occur in rotating clouds, where the rotation velocity can be understood as an external control parameter. Experimentally, the rotation velocity can be controlled via the misalignment of the trapping beams and their intensity.

\par
In conclusion, we have developed a kinetic theory for elementary excitations - hydbrid phonons -  carrying orbital angular momentum in large magneto optical traps, taking into consideration the collective effects emerging from the exchange of scattered photons. The resulting kinetic dispersion relation, formally similar to that of a plane wave, was shown to depend on an effective wave vector, $q_{\rm eff}$, in which the helical structure of the wavefront is embedded. Explicit expressions for the dispersion relation and kinetic damping rates were derived for a cold gas in thermal equilibrium. The contribution from the orbital angular momentum states was encoded into the dimensionless paraxial parameter $\eta = k/ l q_\theta$, which is basically the ratio between the longitudinal and azimuthal wave vectors, $k$, and $q_\theta$, respectively. By setting $\eta \rightarrow \infty$ we recovered the expected result for a plane density wave. 
\par
The case of a rotating cloud of atoms in a ring-shaped structure was considered, in which a new poloidal velocity distribution function was introduced and justified based on the typical experimental scales for the system temperature, radius and angular velocity. A positive damping term from the poloidal velocity distribution can be large enough to compensate for the negative axial damping, in realistic experimental scenarios. In this way, vortex phonons with a finite orbital angular momentum were shown to cross the instability threshold while the plane waves remain kinetically damped. This opens the possibility of easily detecting this vortex instability in rotating clouds of cold atoms in large magneto optical traps. Note that the instability mechanism resembles the ``bump-on-tail" instabilities usually found in plasma physics scenarios, with non-equilibrium velocity distributions where regions of positive slope, $\partial W_0 / \partial v > 0$, act as free energy source for the waves, feeding the instability. In our case, the ``bumpy-tail" velocity distribution emerges from the rotation of the cloud and, by scanning through the $\beta$ parameter, a kinetic instability should be triggered and easily detected through standard fluorescence, or absorption imaging techniques. As a final remark we note that the ring-shaped structure is not mandatory for the instability to occur, as the essential ingredient is always a rotation and a ``bump" in the poloidal velocity distribution, where angular momentum from this region is transferred to the twisted wave. In order to observe the onset of instability in such a situation, a filtered image of the fluorescence of the outer shell should be recorded. A real-time analysis should therefore contain oscillations in the fluorescence signal, at the frequency of the most unstable mode. The measurement of the most unstable mode can then be performed by taking multiple realizations of the Fourier transform of the fluorescence signal. 

\par
JR acknowledges the financial support of FCT - Funda\c{c}\~{a}o da Ci\^{e}ncia e Tecnologia through the grant number SFRH/BD/52323/2013 and HT acknowledges the financial support of the ERC synergy grant UQUAM.

\bibliography{ref}

\end{document}